\documentclass[twocolumn,english,final,journal]{IEEEtran}
\usepackage[T1]{fontenc}
\usepackage{babel}
\usepackage{array}
\usepackage{multirow}
\usepackage{amsthm}
\usepackage{amsmath}
\usepackage{graphicx}
\usepackage{epstopdf}
\usepackage{cite}

\usepackage[unicode=true,
 bookmarks=true,bookmarksnumbered=true,bookmarksopen=true,bookmarksopenlevel=1,
 breaklinks=false,pdfborder={0 0 0},backref=false,colorlinks=false]
 {hyperref}
\hypersetup{pdftitle={Your Title},
 pdfauthor={Your Name},
 pdfpagelayout=OneColumn, pdfnewwindow=true, pdfstartview=XYZ, plainpages=false}
\usepackage{breakurl}

\usepackage{enumitem}

\makeatletter

 \let\oldforeign@language\foreign@language
 \DeclareRobustCommand{\foreign@language}[1]{%
   \lowercase{\oldforeign@language{#1}}}
\theoremstyle{plain}
\newtheorem{thm}{\protect\theoremname}
\theoremstyle{definition}
\newtheorem{defn}[thm]{\protect\definitionname}

\ifCLASSOPTIONcompsoc
\usepackage[caption=false,font=normalsize,labelfont=sf,textfont=sf]{subfig}
\usepackage[compress]{cite}
\usepackage{algorithm}
\usepackage{algorithmic}
\usepackage{supertabular}

\else
\usepackage[caption=false,font=footnotesize]{subfig}
\usepackage{cite}
\usepackage{algorithm}
\usepackage{algorithmic}

\fi

\@ifundefined{showcaptionsetup}{}{%
 \PassOptionsToPackage{caption=false}{subfig}}
\usepackage{subfig}
\makeatother

\providecommand{\definitionname}{Definition}
\providecommand{\theoremname}{Theorem}

\allowdisplaybreaks

\title{Vulnerability Analysis and Consequences of False Data Injection Attack
on Power System State Estimation}

\author{Jingwen Liang,~\IEEEmembership{Student Member,~IEEE,} Lalitha Sankar,~\IEEEmembership{Member, IEEE,}
and~Oliver Kosut,~\IEEEmembership{Member, IEEE}%
\thanks{J. Liang, L. Sankar, and O. Kosut are with the Department of Electrical,
Computer and Energy Engineering, Arizona State University, Tempe,
AZ 85281 USA, e-mail: \protect\href{mailto:jliang19@asu.edu}{jliang19@asu.edu},
\protect\href{mailto:lsankar@asu.edu}{lsankar@asu.edu}, and \protect\href{mailto:okosut@asu.edu}{okosut@asu.edu}.
 This work is supported in part by NSF grant CPS-1449080.}}

\begin{document}

\maketitle
\IEEEpeerreviewmaketitle
\begin{abstract}
An unobservable false data injection (FDI) attack on AC state estimation
(SE) is introduced and its consequences on the physical system are
studied. With a focus on understanding the physical consequences of
FDI attacks, a bi-level optimization problem is introduced whose objective
is to maximize the physical line flows subsequent to an FDI attack
on DC SE. The maximization is subject to constraints on both attacker
resources (size of attack) and attack detection (limiting load shifts)
as well as those required by DC optimal power flow (OPF) following
SE. The resulting attacks are tested on a more realistic non-linear
system model using AC state estimation and ACOPF, and it is shown
that, with an appropriately chosen sub-network, the attacker can overload
transmission lines with moderate shifts of load. \end{abstract}
\begin{IEEEkeywords}
false data injection, state estimation, optimization, vulnerability analysis.
\end{IEEEkeywords}
\section{Introduction}

\IEEEPARstart{W}{ith} the increasing integration of real-time
monitoring, sensing, control, and communication, the electric power systems are becoming increasingly efficient and controllable.
However, the tight integration also makes the system more vulnerable
to cyber attacks with potentially serious physical consequences. Therefore,
assessment and evaluation of possible attacks and consequences before
an actual attack happens is extremely instructive to the utilities:
procedures for detecting potential attack incidents is an
important supplements to the secure operation of the power system.

There is much interest in studying cyber attacks on the electric power
system. This includes attacks on system states \citen{Liu09, Teixeira2010,Kosut11, Pasqualetti2011, Sandberg10,Rahman13},
system topology \citen{KIM13TOPO, Kim13UNKNOWN}, generator
dynamics \citen{ Kundur2010}, and energy markets
\citen{Xie2011, Giani12, Jia2014}. While several
classes of cyber-attacks have been identified, consequences of such
attacks on the electric power system are less understood. It is this
aspect that we focus on in this paper. To this end, we introduce a
class of false data injection (FDI) attacks on AC state estimation
(SE) designed to cause physical line overflows. In fact, such an attack
can potentially lead to cascading failures since a sustained attack
can ensure that the physical line overflow is not detected through
the cyber measurements.

\subsection{Contributions}

The contributions of this paper are two-fold. First, we introduce
a sophisticated unobservable attack on AC state estimation that takes
into account the sequential data processing functionalities in the
cyber layer of the electric power system (see Fig. \ref{fig:Temporal-nature-processing}). Our attack
models a sophisticated attacker with access to measurements in a small
sub-network and with the intention of creating significant changes
to the physical network that can have potentially damaging consequences,
if undetected. Enabling physical consequences requires the attacker to change measurements that leads to redispatch,
and eventually, line overloads. To this end, we formulate a bi-level attack
optimization problem with the objective of causing a physical line
overflow via an unobservable attack on AC state estimation subject
to constraints on: (i) number of meters to attack (limited resources
constraint) and (ii) load shifts (to limit operator detection).
Since a line overflow requires modeling the system level redispatch
\textit{subsequent to the attack}, our optimization problem has embedded
in it a second level redispatch optimization problem. The second contribution
of our work is to highlight the consequences of our proposed attack
on a non-linear system model with AC SE and ACOPF. We use the optimal
attack vector obtained from our optimization problem to do so. We
show that our attack model can successfully lead to line overflows
for an RTS-24-bus system with moderate load shifts and attack sizes.

\subsection{State of the Art}

FDI attacks have gained much interest in the literature starting from
Liu \textit{et al.}'s work on unobservable attacks on DC SE \citen{Liu09}.
Their work shows that an attacker can change the system state without being detected by the bad data detection algorithm
within SE if the attack vector is chosen judiciously to mimic typical
SCADA measurements. Kosut \textit{et al.} discuss the trade-off between
maximizing estimation error at the control center and minimizing detection
probability of the attack \citen{Kosut11}. 

For attacks  restricted to a sub-network of the system network graph, the authors in \citen{Hug2012} introduce an algorithm
to determine an \textit{attack subgraph} and show that such a sub-network
must be bounded by buses with injections. Furthermore, the authors
also show that a sophisticated attack using AC SE requires the attacker
to estimate the system states for its subgraph. Recently, in \citen{Liang14}, we build upon \citen{Hug2012} to introduce an AC attack restricted to a subgraph and show that it suffice for the attacker to perform local SE to launch an unobservable attack.

In this paper we extend \citen{Liang14} to study attack consequence. To this end, we use an attack subgraph
and determine the optimal attack via a bi-level optimization problem.
Bi-level attack optimization problems in the context of attacks are
considered in \citen{Salmeron04,Yuan11,YUAN12}.
In all cases, the optimization problems include both the attacker's
goal (unobservable attack on DC SE) as well as the ensuing system
response (OPF), leading to a bi-level optimization problem. However, the goal of the optimization in the aforementioned
papers is to increase the operating costs for the system. While costs
are relevant to the electric power system operation, cyber attacks
with physical consequences can be more damaging.

The optimization problems in \citen{Yuan11} and \citen{YUAN12} take
into account the fact that FDI attacks lead to an inevitable load
shift at the buses in the subgraph and include a constraint on the load shift magnitude to limit detection. In this
paper, we take this a step further and restrict not only the load
shift magnitude but also the size of attack subgraph to simultaneously
model the observability and limited resources constraints.

The remainder of this paper is organized as follows. Sec. \ref{sec:Problem-formulation:-system}
introduces the general system and attack model. Sec. \ref{sec:Attack-strategy}
discusses the different attack strategies for unobservable attacks.
Sec. \ref{sec:optimization-problem} presents a bi-level optimization
formulation to identify the worst-case overflow attack. Sec. \ref{sec:Simulation-Results}
presents and analyzes the numerical results for a test system. Sec.
\ref{sec:Conclusions-and-future} draws the conclusion of this paper
and presents the direction of future works.

\section{Problem formulation: system and attack model\label{sec:Problem-formulation:-system}}

\subsection{Temporal nature processing of the grid}

\begin{figure*}[tbh]
\centering{}
\includegraphics[bb=15bp 26bp 770bp 427bp,clip,scale=0.65]{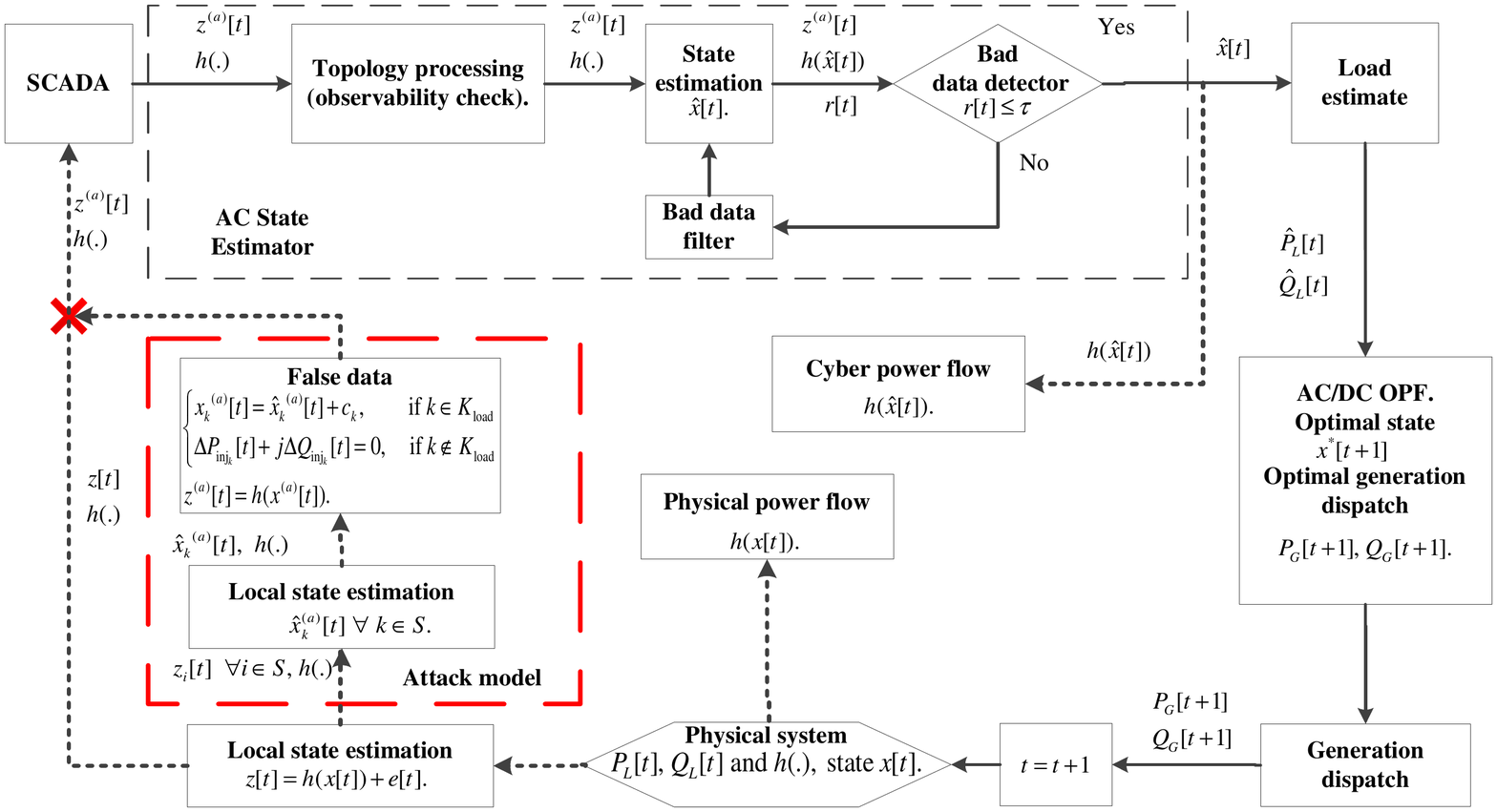}
\caption{Illustration of temporal processing of the grid and attack model.
\label{fig:Temporal-nature-processing}}
\end{figure*}

Fig. \ref{fig:Temporal-nature-processing} illustrates the temporal
nature of processing in the grid and the attack model. Assume
a system with $n_{b}$ buses, $n_{br}$ branches, and $n_{g}$ generators.
Active and reactive load of each buses are represented by $P_{L}$
and $Q_{L}$, respectively. Measurement and estimated measurement
residue are denoted as $z$ and $r$, respectively. In the bad data
detector, $\tau$ is the residue threshold and $x=[V\mbox{, }\theta]^{T}$
is the system state, where $V$ is bus voltage magnitude and $\theta$
is bus voltage angle. The function $h(\cdot)$ denotes the non-linear
function that gives the measurements. This function depends only on the system
topology. Estimated values are denoted with a hat, e.g. $\hat{x},\hat{V},\hat{\theta}$.

As shown in Fig. \ref{fig:Temporal-nature-processing}, generation
dispatch control decisions made at the control center depend on the
noisy measurements provided by the SCADA system. If these measurements
are corrupted by an attacker and pass the bad data detector, 
they can directly influence the control decisions for the next time
interval. Since the process occurs in the same manner for each time
$t$, we drop the functional dependence on $t$ for the rest of this section.
The major blocks shown in Fig. \ref{fig:Temporal-nature-processing}
are discussed in detailed in the following subsections.

\subsection{Measurements}
The AC measurement model follows the non-linear relationship
\begin{equation}
z=h(x)+e\mbox{.}\label{Measurement_model}
\end{equation}
where $z$, $e$ and $x$ are $m\times1$, $m\times1$ and $n\times1$
vectors with entries $z_{i}$, $e_{i}$ and $x_{k}$, respectively
$i\in\{1,\ldots,m\}$ and $k\in\{1,\ldots,n\}$. $z_{i}$ is the $i^{\mbox{th}}$
measurement of the system: line power flows, bus voltage and line
current magnitude, etc. $e_{i}$ is the $i^{\mbox{\textrm{th}}}$
measurement error, assuming to be independent and Gaussian distributed
with $0$ mean and $\sigma_{i}^{2}$ covariance.

\subsection{State estimation \label{sub:State-estimation}}

As illustrated in Fig. \ref{fig:Temporal-nature-processing}, all
raw measurements are first passed through an observability check.
If there are enough measurements, the system will be observable; otherwise,
the system is divided into several observable islands.

State estimation is used to determine the most likely state of the
system given the available noisy measurements. In AC state estimation,
the state variables are solved as a least square problem with an objective
function \citen{Abur00}
\begin{equation}
\hat{x}=\mbox{arg}\min J(x)=(h(x)-z)^{T}R^{-1}(h(x)-z)\label{eq:LSP_obj}
\end{equation}
where $R=\mbox{diag}\{\sigma_{1}^{2},\sigma_{2}^{2},\ldots,\sigma_{m}^{2}\}$
and $\hat{x}=[\hat{V}\mbox{, }\hat{\theta}]^{T}$ is the estimated
state.

Subsequent to SE, the bad data detector filters noisy measurement and guarantees the
accuracy of estimation, using $\chi^{2}$ test.
\subsection{AC and DC optimal power flow}

The AC optimal power flow (OPF) takes the following form:
\begin{align}
\underset{x}{\mbox{minimize}}\:\:\; & f(P_{G})\nonumber \\
\mbox{subject to}\:\; & G(x)=0\label{eq:OPF_eq}\\
 & F(x)\leq0\label{eq:OPF_ineq}\\
 & x_{\max}\leq x\leq x_{\max}\label{eq:var_ValueRange-AC}
\end{align}
 where $f(.)$ is the generation cost function and $x=[V\mbox{, }\theta\mbox{, }P_{G}\mbox{, }Q_{G}]^{T}$ is the  variable of the optimization problem. Inequality constraint,
i.e. the line thermal limits, is denoted as $F$ and equality constraint,
i.e. the node power balance is denoted as $G$. Both $F$ and $G$
are non-linear constraints, since there are active and reactive power
involved.

DCOPF approximates $G$ and $F$ around $V=1$, $\theta=0$ by their first order Taylor
expansion:
\begin{align}
\mbox{\ensuremath{\underset{\theta,\mbox{\ensuremath{P_{G}}}}{\mbox{minimize}}}\:\:}\; & f(P_{G})\nonumber \\
\mbox{subject to}\:\; & -H_{1}\theta+P_{G}-P_{L}=0\label{eq:eq_DCOPF}\\
 & -P^{\max}\leq H_{2}\theta\leq P^{\max}\label{eq:ineq_DCOPF}\\
 & P_{G}^{\min}\leq P_{G}\leq P_{G}^{\max}.\label{eq:var_ValueRange-DC}
\end{align}
where
\begin{itemize}
\item $H_{1}$ is the matrix of dependencies between power injection  and
state $\theta$.
\item $H_{2}$ is the matrix of dependencies between branch power flow and
state $\theta$.
\item $P^{\max}$ is the thermal limit.
\item $P_{G}^{\min}$and $P_{G}^{\max}$ are the generator capacity lower
and upper limit, respectively.
\end{itemize}

\subsection{Attack model\label{sub:attack-model}}

We first assume that the attacker has following capabilities:
\begin{enumerate} 
\item The attacker has access to all measurements and topology information of a small area $\mathcal{S}$ bounded by buses. The set of all measurement indices in $\mathcal{S}$ is denoted as $\mathcal{I_{S}}$ and the set of all state indices in $\mathcal{S}$ is denoted as $\mathcal{K_{S}}$. 
\item The attacker can change or replace all measurements in $\mathcal{S}$. 
\item The attacker has computational capability.  
\end{enumerate}

As discussed in \citen{Liang14}, according to \eqref{Measurement_model}, suppose the $i^{\mbox{th}}$ measurement prior to attack is $z_{i}=h_{i}(x)+e_{i}$, the general attack model changes the $i^{\mbox{th}}$ measurement $z_{i}$ to $z_{i}^{(a)}$ such that
\begin{equation}
z_{i}^{(a)}=\begin{cases}
\begin{array}{c}
z_{i}\\
\tilde{z}_{i}
\end{array} & \begin{array}{c}
\mathrm{if}\, i\notin\mathcal{I}_{\mathcal{S}}\\
\mathrm{if}\, i\in\mathcal{I}_{\mathcal{S}}
\end{array}\end{cases}\label{eq:aGeneral_attack_model}
\end{equation}
where $\tilde{z}_{i}$ is chosen by attacker.

\section{Attack strategy\label{sec:Attack-strategy}}

\subsection{Unobservable attack \label{sub:Unobservable-attack}}
\begin{defn}
An attack is \emph{unobservable} for a measurement model $h(\cdot)$
if, in the absence of measurement noise, there exists a $c\neq0$
such that $z_{i}^{(a)}=h_{i}(x+c)$ for all $i$.

Therefore, for the attacker to execute an unobservable attack, again
assuming no measurement noise, \eqref{eq:aGeneral_attack_model} becomes
\begin{equation}
z_{i}^{(a)}=\begin{cases}
\begin{array}{c}
z_{i}\qquad\quad\quad\\
h_{i}(x+c)
\end{array} & \begin{array}{c}
\mathrm{if}\, i\notin\mathcal{I_{S}}\\
\mathrm{if}\, i\in\mathcal{I_{S}}.
\end{array}\end{cases}\label{eq:Unobe_attack}
\end{equation}

\end{defn}
From \eqref{eq:Unobe_attack}, if the $k^{\mbox{th}}$ state $x_{k}$
is required to compute $h_{i}(x)$ for any $i\notin\mathcal{I_{S}}$,
then for any unobservable attack the corresponding $k^{\mbox{th}}$
entry in attack vector must satisfy $c_{k}=0$. That is, for an attack
region $\text{\ensuremath{\mathcal{S}}}$, not all the bus states
in it can be changed. The attack region must be bounded by a set of
buses without state changes however with measurement changes. To identify
such a collection of one or more buses in $\mathcal{S}$, we first
distinguish between two types of buses based on the presence of load.
We henceforth identify buses with load as \emph{load buses}. $\mathcal{K}_{\textrm{load}}$
denotes the bus indices of load bus. An attacker can attack either
type of bus. However, since the injections of non-load buses are known
to the control center, attacking a non-load bus implies that the measurements
at the closest load buses also need to be changed to ensure that the
nodal power balance is maintained. In \citen{Hug2012}, a method is
introduced to identify a subgraph of the network that allows an attacker
to perform an unobservable attack. We use a similar method, as summarized
as follow. Let $k$ be a target load bus, the corresponding \emph{single-target-bus
attack subgraph} $\mathcal{S}_{k}$ is constructed by following steps:
\begin{enumerate}
\item Include bus $k$ in $\mathcal{S}_{k}$. 
\item Extend $\mathcal{S}_{k}$ from bus $k$ by including all buses and
branches that are connected to bus $k$.
\item \label{enu:If-there-is}If there is a non-load bus on the boundary
of $\mathcal{S}_{k}$, extend $\mathcal{S}_{k}$ to include all adjacent
buses of such a boundary bus.
\item Repeat \eqref{enu:If-there-is} until all buses on the boundary are
load buses or $\mathcal{S}_{k}$ can not be extended anymore.
\end{enumerate}
\begin{figure}
\includegraphics[bb=30bp 17bp 550bp 132bp,clip,scale=0.45]{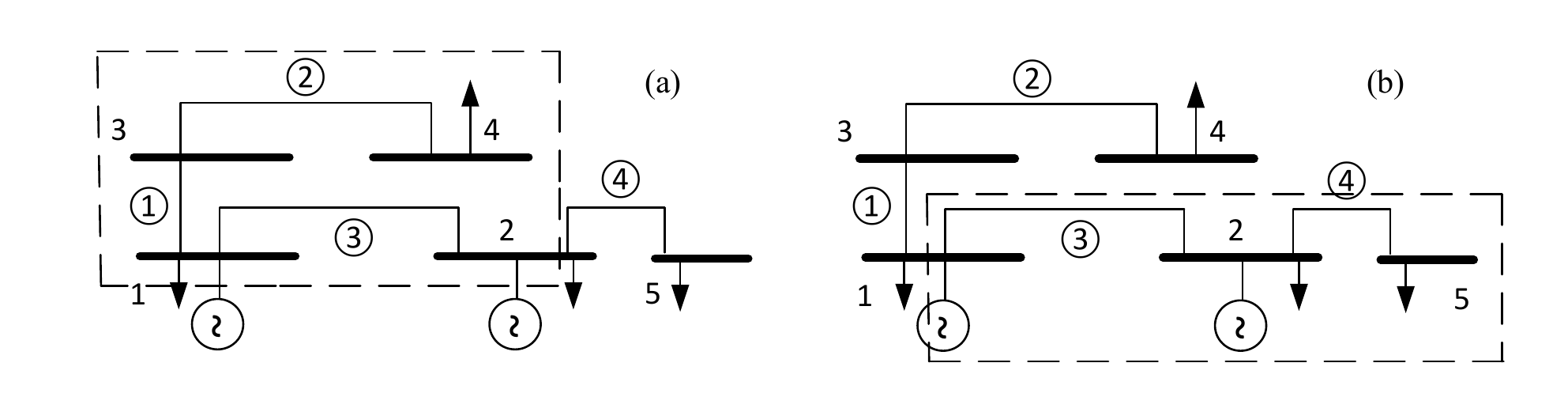}
\caption{Examples of single-target-bus attack subgraph. Fig. 2a shows the subgraph
with target bus 1 and Fig. 2b shows the subgraph with target bus 2.}
\label{fig:Figure 1} 
\end{figure}
The steps above give an attack subgraph that includes the target load
bus and is bounded by load buses. Fig. \ref{fig:Figure 1} shows two
simple examples of single-target-bus attack subgraphs. The choice
of the final attack subgraph $\mathcal{S}$, however, can be a union
of several single-target-bus attack subgraphs:
\begin{equation}
\mathcal{S}=\underset{k\mbox{ : }c_{k}\neq0\cap k\in\mathcal{K}_{\textrm{load}}}{\bigcup}\mathcal{S}_{k}\mbox{.}
\end{equation}
This choice of attack subgraph results in estimated load changes at
all load bus within $\mathcal{S}$ while no net load changes in the
system. 

\subsection{DC attack \label{sub:DC-attack}}

Since \eqref{eq:Unobe_attack} is nonlinear and generally hard to
solve, it is reasonable for the attacker to first consider a simplified
DC attack. As \citen{Liu09} demonstrated, by knowing system Jacobian
matrix $H$, an attacker can intelligently construct an unobservable
attack vector $a=Hc$ such that $z_{i}^{(a)}=z_{i}+a.$

Thus, \eqref{eq:Unobe_attack} becomes
\begin{equation}
z_{i}^{(a)}=\begin{cases}
\begin{array}{c}
z_{i}\qquad\;\quad\;\qquad\\
z_{i}+H_{(i,:)}c
\end{array} & \begin{array}{c}
\mbox{if }i\notin\mathcal{I}_{\mathcal{S}P}\\
\mbox{if }i\in\mathcal{I}_{\mathcal{S}P}
\end{array}\end{cases}\label{eq:DC_attack}
\end{equation}
where, $\mathcal{I}_{P}$ denotes the set of indices of active power
measurements, $\mathcal{I}_{\mathcal{S}P}=\mathcal{I}_{\mathcal{S}}\cap\mathcal{I}_{P}$,
and $H_{(i,:)}$ denotes the $i^{\mbox{th}}$ row of $H$.

Though DC attack is easy to construct, it is not an unobservable attack
for AC state estimator. Without taking reactive power flow into account,
a DC attack will be detected when $c$ is too large.

\subsection{AC attack \label{sub:AC-attack}}

From \eqref{eq:Unobe_attack}, in contrary to DC attack, it seems
that the attacker must know all the state values that appear in $h_{i}(.)$,
for all $i\in I_{\mathcal{S}}$, to construct $z_{i}$ precisely.
However, this information is not available to the attacker. Thus,
attacker can use the following steps to construct $z_{i}^{(a)}$:
\begin{enumerate}
\item The attacker first chooses the non-zero entries in $c$ only for the
load buses. These non-zero entries correspond to the center buses
for the attack subgraph.
\item Use the protocol in Sec. \ref{sub:Unobservable-attack} and choose
$\mathcal{S}$ for the desired attack.
\item Given the measurements that are available to the attacker in $\mathcal{S}$,
perform local AC state estimation to find $\hat{x}_{k}^{(a)}$. The
slack bus may be chosen arbitrarily among all load buses.
\item For all load buses $k$, set $x_{k}^{(a)}=\hat{x}_{k}^{(a)}+c_{k}$.
\item Since the injection of non-load buses can not be changed, the states
of non-load buses are dependent on the state of all the buses that
connected to them. Therefore, the attacker has the nodal balance equation
for each non-load bus $k$ in $\mathcal{S}$: 
\begin{align}
P_{\textrm{inj}_{k}}=V_{k}\sum_{i\in\mathcal{N_{\mathnormal{k}}}}V_{i}(G_{ki}\mbox{cos}\theta_{ki}+B_{ki}\mbox{sin}\theta_{ki})\\
Q_{\textrm{inj}_{k}}=V_{k}\sum_{i\in\mathcal{N_{\mathnormal{k}}}}V_{i}(G_{ki}\mbox{sin}\theta_{ki}-B_{ki}\mbox{cos}\theta_{ki})
\end{align}
where $G_{ki}+jB_{ki}$ is the $(k,i)^{\textrm{th}}$ entry of the
complex bus admittance matrix and $\theta_{ki}=\theta_{k}-\theta_{i}$
is the angle difference between bus $k$ and $i$. These equations
can be solved by iterative methods such as Newton-Raphson method.

\item With all the computed state information, the attacker can therefore
compute the false measurements $z^{(a)}$ such that
\end{enumerate}
\begin{equation}
z_{i}^{(a)}=\begin{cases}
\begin{array}{c}
z_{i}\qquad\;\quad\:\;\\
h_{i}(x^{(a)})\quad\quad
\end{array} & \begin{array}{c}
\mathrm{\mbox{if }}i\notin\mathcal{I}_{\mathcal{S}}\\
\mbox{\ensuremath{\mathrm{\mbox{if }}}}i\in\mathcal{I}_{\mathcal{S}}.
\end{array}\end{cases}\label{eq:AC_attack}
\end{equation}

\section{Optimization problem for the worst-case line overload attack\label{sec:optimization-problem}}

The aim of the unobservable attack is to maximize the physical line
flow for a chosen line in the attack subgraph. However, the attacker,
in general, has limited resources to change states; furthermore, the
attacker would also like to design the attack to avoid detection over
the various computing units in EMS. This leads to a constrained optimization
problem. Specifically, we model the two conflicting goals of the attacker
as follows: the limited resource constraint is modeled by a sparsity
constraint in which we limit the number of center buses at which states
can be changed. The detectability constraint is modeled by limiting
the cyber load shifts that result from the FDI attacks. This is because
a large deviation in estimated load from normal operational values
will be detected as an anomalous event by the operators. The sparsity
constraint capturing the limited resource is modeled as an $l_{0}$-norm
constraint. This is, in general, intractable, and therefore, we relax
it to an $l_{1}$-norm constraint. In addition to the two constraints,
since the physical line flow is a consequence of the control center
re-dispatch generation, the attack optimization process has to include
the OPF subsequent to state estimation as a sub-problem. The resulting
problem is a bi-level optimization problem.

Generally, an optimal dispatch can be the result of different load
patterns. As a result, there are numerous solutions of attack vector
that lead to the same physical line flow on the target line. Among
these, the goal of the optimization is to choose the one with the
smallest $l_{1}$-norm, and hence, $l_{0}$-norm to satisfy the limited
resources constraint. This, in turn, requires a second entry in the
objective function where we determine the sparest attack vector among
the same maximal power flow on the target branch.

The attacker's influences over the system can be formulated as an
optimization problem (with attacker's objective) embedded with a sub-problem
(with operator\textquoteright{}s objective). Similar to the authors
in \citen{Salmeron04,Yuan11}, we model the optimal attack problem
as a bi-level optimization problem with an objective to maximize the
power flow on branch $l$ while to change as few states as possible:
\begin{align}
\negthickspace\negthickspace\negthickspace\mbox{maximize}\:\: & \qquad\:\:\:\:\: P_{l}-\mbox{\ensuremath{\gamma\left\Vert c\right\Vert _{0}}}\label{eq:Obj}\\
\mbox{\negthickspace\negthickspace\negthickspace subject to }\nonumber \\
 & \negthickspace\negthickspace\negthickspace\negthickspace\negthickspace\negthickspace\negthickspace\negthickspace\negthickspace\negthickspace\negthickspace\negthickspace\negthickspace P=H_{2}(\theta^{\star}-c)\label{eq:PF_EQ}\\
 & \negthickspace\negthickspace\negthickspace\negthickspace\negthickspace\negthickspace\negthickspace\negthickspace\negthickspace\negthickspace\negthickspace\negthickspace\negthickspace-L_{S}P\leq H_{1}c\leq L_{S}P_{L}\label{eq:LS_Constra_1}\\
 & \negthickspace\negthickspace\negthickspace\negthickspace\negthickspace\negthickspace\negthickspace\negthickspace\negthickspace\negthickspace\negthickspace\negthickspace\negthickspace\left\Vert c\right\Vert _{0}\leq N_{0}\label{eq:L0Norm_Constr_1}\\
 & \negthickspace\negthickspace\negthickspace\negthickspace\negthickspace\negthickspace\negthickspace\negthickspace\negthickspace\negthickspace\negthickspace\negthickspace\negthickspace\left\{ \mathbf{\theta^{\star}},P_{G}^{\star},R^{\star}\right\} =\arg\left\{ \min_{\theta,P_{G},R}{\displaystyle \sum_{g=1}^{n_{g}}f_{g}(P_{G_{g}})}+\sum_{l=1}^{n_{br}}\mathcal{P}_{l}(R_{l})\right\} \label{eq:DCOPF1}\\
 & \begin{array}{cc}
\negthickspace\negthickspace\negthickspace\negthickspace\negthickspace\negthickspace\negthickspace\negthickspace\negthickspace\negthickspace\negthickspace\negthickspace{P_{G}-H_{1}(\theta-c)-P_{L}=0} & {(\mathbf{\upsilon})}\end{array}\label{eq:DCOPF2}\\
 & \begin{array}{cc}
\negthickspace\negthickspace\negthickspace\negthickspace\negthickspace\negthickspace\negthickspace\negthickspace\negthickspace\negthickspace\negthickspace\negthickspace{-P^{\max}-R\leq H_{2}\mathbf{\theta}\leq P^{\ensuremath{\max}}+R\:} & {(\lambda^{+},\lambda^{-})}\end{array}\label{eq:DCOPF3}\\
 & \begin{array}{cc}
\negthickspace\negthickspace\negthickspace\negthickspace\negthickspace\negthickspace\negthickspace\negthickspace\negthickspace\negthickspace\negthickspace\negthickspace{P_{G}^{\min}\leq P_{G}\leq P_{G}^{\max}} & {(\alpha^{+},\alpha^{-})}\end{array}\label{eq:DCOPF4}\\
 & \begin{array}{cc}
\negthickspace\negthickspace\negthickspace\negthickspace\negthickspace\negthickspace\negthickspace\negthickspace\negthickspace\negthickspace\negthickspace\negthickspace{0\leq R} & {(\beta)}\end{array}\label{eq:DCOPF5}
\end{align}
where the variables:
\begin{description}[leftmargin=1.8cm,style=multiline]
\item[$P$] is the $n_{br}\times1$ vector of branch power flow;
\item[$c$] is the $n_{b}\times1$ attack vector;
\item[$\theta,\theta^{\star}$] are $n_{b}\times1$ state variable vectors and optimal variable solved
by DCOPF, respectively;
\item[$P_{G},P_{G}^{\star}$] are $n_{g}\times1$ vectors of generation dispatch variable and optimal
generation dispatch solved by DCOPF, respectively;
\item[$R,R^{\star}$] are $n_{br}\times1$ vectors of the line relaxation variable, and
optimal line relaxation solved by DCOPF, respectively;
\item[$\upsilon$] is the $n_{b}\times1$ dual variable vector for all equal constraints
in DCOPF;
\item[$\lambda^{+},\lambda^{-}$] are $n_{br}\times1$ dual variable vectors of the upper and lower
bound of thermal limits, respectively;
\item[$\alpha^{+},\alpha^{-}$] are $n_{g}\times1$ dual variable vectors of the upper and lower bound
of generator capacity, respectively;
\end{description}
and the parameters:
\begin{description}[leftmargin=1.8cm,style=multiline]
\item[$L_{S}$] is the load shift factor;
\item[$P_{L}$] is the $n_{b}\times1$ vector of active load at each bus;
\item[$N_{0}$] is the $l_{0}$-norm constraint integer;
\item[$H_{1}$] is the $n_{b}\times n_{b}$ matrix of dependencies between power injection
measurements and state variables;
\item[$H_{2}$] is the $n_{nb}\times n_{b}$ matrix of dependencies between power
flow measurements and state variables;
\item[$f_{g}$] is the cost function of the $g^{\mbox{th}}$ generator;
\item[$\mathcal{P}_{l}$]  is the penalty function of relaxing the $l^{\mbox{th}}$ line;
\item[$P^{\max}$] is the $n_{br}\times1$ vector of line thermal limit;
\item[$P_{G}^{\min},P_{G}^{\max}$] are $n_{g}\times1$ vectors of minimum and maximum generator output,
respectively;
\item[$\gamma$] the weight of the norm of attack vector $c$.
\end{description}

We define $l_{0}$-norm as appropriate quantities summed over
only the load buses. Thus, the $l_{0}$-norm,$\left\Vert c\right\Vert _{0}$,
of the attack vector $c$ is defined as
\begin{equation}
\left\Vert c\right\Vert _{0}=\sum_{k\in\mathcal{K}_{\textrm{load}}}^{n_{b}}1(c_{k}\neq0).\label{eq:N0_computation}
\end{equation}

Recall the goal of optimization is to maximize $P_{l}$ while finding
the sparsest attack among all the possible attack vector. Thus, due
to the trade-off between the maximum power flow and the corresponding
sparest attack vector, thus the optimization objective is $P_{l}-\mbox{\ensuremath{\gamma\left\Vert c\right\Vert _{0}}}$.
The weight $\gamma$ is chosen to be a small and positive value such
it in general contributes minimal to the objective. Note that \eqref{eq:PF_EQ}--\eqref{eq:L0Norm_Constr_1}
are the attack related constraints. The constraints in \eqref{eq:PF_EQ}
model the unobservability of the attack and the constraints in \eqref{eq:LS_Constra_1}--\eqref{eq:L0Norm_Constr_1}
model the attacker's limited ability: the attacker can alter up to
$N_{0}$ states (not necessarily alter all of them) and the resulting
change in load shift is limited to $L_{S}P_{L}$. A standard DCOPF
with a thermal limit relaxation penalty is modeled by \eqref{eq:DCOPF1}--\eqref{eq:DCOPF5}.
The penalty function in \eqref{eq:DCOPF1} ensures the second level
OPF converge thus the first level problem to return a solution.

Since \eqref{eq:L0Norm_Constr_1} is a modified $l_{0}$-norm constraint,
it is a complex non-linear constraint and generally non-convex. In
this paper, we relax it to a corresponding $l_{1}$-norm constraint
as
\begin{equation}
\left\Vert c\right\Vert _{1}=\sum_{k\in\mathcal{K}_{\textrm{load}}}\left|c_{k}\right|\leq N_{1}\label{eq:L1Norm_Constr_1}
\end{equation}
where $N_{1}$ is non-negative. Since \eqref{eq:L1Norm_Constr_1}
is a non-linear constraint and we rewrite it as
\begin{align}
-c_{k} & \leq s_{k},  &  c_{k} & \leq s_{k}, & \sum_{k\in\mathcal{K}_{\textrm{load}}}s_{k}\leq N_{1}.\label{eq:N1}
\end{align}
where $s$ is a slack variable.

For the embedded OPF problem, the optimal solution can be found at
the point which satisfies the KKT optimality condition with zero duality
gap since it is a convex optimization problem \citen{CV}. We use
this fact to further replace the embedded DCOPF problem in \eqref{eq:DCOPF1}
with its KKT conditions below, along with \eqref{eq:DCOPF2}--\eqref{eq:DCOPF5},
as 
\begin{gather}
\left[\lambda^{+};\lambda^{-};\alpha^{+};\alpha^{-};\beta\right]\geq0\label{eq:Dual>0}\\
\mbox{diag}\left(\left[\lambda^{+};\lambda^{-}\right]\right)\left(\left[H_{2};-H_{2}\right]\theta^{\star}-\left[P^{\max}+R^{\star}\right]\left[I;-I\right]\right)=0\label{eq:ComplementarySlack1}\\
\mbox{diag}\left(\left[\alpha^{+};\alpha^{-}\right]\right)\left(\left[I;-I\right]P_{G}^{\star}-\left[P_{G}^{\max};-P_{G}^{\min}\right]\right)=0\\
-\mbox{diag}(\beta)R^{\star}=0\label{eq:ComplementarySlack2}\\
\nabla({\displaystyle \sum_{g=1}^{n_{g}}f_{g}(P_{G_{g}}^{\star})}+{\displaystyle \sum_{l=1}^{n_{br}}\mathcal{P}_{l}(R_{l}^{\star})})\nonumber \\
+\left[\lambda^{+};\lambda^{-}\right]^{T}\nabla\left(\left[H_{2};-H_{2}\right]\theta^{\star}-\left[P^{\max}+R^{\star}\right]\left[I;-I\right]\right)\nonumber \\
+\left[\alpha^{+};\alpha^{-}\right]^{T}\nabla\left(\left[I;-I\right]P_{G}^{\star}-\left[P_{G}^{\max};-P_{G}^{\min}\right]\right)\label{eq:PartialGradient}\\
-\beta^{T}\nabla R^{\star}+\upsilon^{T}\nabla[P_{G}^{\star}-H_{1}(\theta^{*}-c)-P_{L}]=0\nonumber 
\end{gather}
where \eqref{eq:ComplementarySlack1}--\eqref{eq:ComplementarySlack2}
are the complementary slackness condition for constraint \eqref{eq:DCOPF3}--\eqref{eq:DCOPF5}
and \eqref{eq:PartialGradient} is the partial gradient optimal condition.
Though \eqref{eq:ComplementarySlack1}--\eqref{eq:ComplementarySlack2}
are non-linear, they have specially distinctive nature. For instance,
the $j^{\textrm{th}}$ equation in \eqref{eq:ComplementarySlack2}
can be separated into two conditions associated with a binary variable
$\delta_{\beta_{j}}$ 
\begin{equation}
\begin{cases}
\beta_{j}\geq0\:\mbox{and \ensuremath{-R_{j}^{\star}=0},} & \mbox{if }\delta_{\beta_{j}}=0\\
\beta_{j}=0\:\mbox{and \ensuremath{-R_{j}^{\star}<0},} & \mbox{if }\delta_{\beta_{j}}=1.
\end{cases}\label{eq:C1}
\end{equation}

In \citen{MacCarl1981}, a procedure is proposed to write \eqref{eq:C1}
as a mixed integer problem given as
\begin{align}
\delta_{\beta_{j}}= \{1,0\},  &\quad\quad  \beta_{j}\leq C\delta_{\beta_{j}}, &  R_{j}^{\star}\leq C(1-\delta_{\beta_{j}}).\label{eq:C2}
\end{align}
If $\delta_{\beta_{j}}=0$, substitute \eqref{eq:DCOPF5} and \eqref{eq:Dual>0}
into \eqref{eq:C2}, we have 
\begin{align}
\delta_{\beta_{j}}=0, & \quad\quad 0\leq\beta_{j}\text{\ensuremath{\leq}}0, &0\leq R_{j}^{\star}\leq C_{j}.\label{eq:R_not_active}
\end{align}
Thus, if $C_{j}$ is large enough to not effect the solution of $R_{j}^{\star}$,
\eqref{eq:R_not_active} is equivalent to the complementary slackness
when the $j^{\textrm{th}}$ constraint in \eqref{eq:DCOPF5} is not
an active constraint. 
Similarly, if $\delta_{\beta_{j}}=1$ and substitute \eqref{eq:DCOPF5}
and \eqref{eq:Dual>0} into \eqref{eq:LinearComplementary2}, we have
\begin{align}
\delta_{\beta_{j}}=1, & \quad\quad 0\leq\beta_{j}\leq C_{j}, & 0\leq R_{j}^{\star}\leq0.\label{eq:R_active}
\end{align}
Again, if $C_{j}$ is large enough to not effect the solution of $\beta_{j}$,
\eqref{eq:R_active} is equivalent to the complementary slackness
when the $j^{\textrm{th}}$ constraint in \eqref{eq:DCOPF5} is an
active constraint. Therefore, \eqref{eq:C2} is equivalent to \eqref{eq:ComplementarySlack2}.

Thus, the whole problem becomes the mixed-integer linear program
\begin{align}
\mbox{maximize}\;\:\qquad & P_{l}-\gamma\mbox{\ensuremath{\sum_{k\in\mathcal{K}_{\textrm{load}}}s_{k}}}\nonumber \\
\mbox{ subject to\qquad\,\ }\nonumber \\
 & \mbox{\negthickspace\negthickspace\negthickspace\negthickspace\negthickspace\negthickspace\negthickspace\negthickspace\eqref{eq:PF_EQ}--\eqref{eq:LS_Constra_1}, \eqref{eq:DCOPF2}--\eqref{eq:DCOPF5}, \eqref{eq:N1}--\mbox{\eqref{eq:Dual>0}, \eqref{eq:PartialGradient}}}\nonumber \\
 & \negthickspace\negthickspace\negthickspace\negthickspace\negthickspace\negthickspace\negthickspace\negthickspace\negthickspace\left\{ \begin{array}{l}
\begin{array}{ccc}
\delta_{\lambda}^{\pm} & \negthickspace\negthickspace\negthickspace= & \negthickspace\negthickspace\negthickspace\{1,0\}\\
\lambda^{\pm} & \negthickspace\negthickspace\negthickspace\leq & \negthickspace\negthickspace\negthickspace C\delta_{\lambda}^{\pm}
\end{array}\\
-H_{2}\theta^{\star}+P^{\max}+R^{\star}\leq C(1-\delta_{\lambda}^{+})\\
+H_{2}\theta^{\star}+P^{\max}+R^{\star}\leq C(1-\delta_{\lambda}^{-})
\end{array}\right.\label{eq:LinearComplementary1}\\
 & \negthickspace\negthickspace\negthickspace\negthickspace\negthickspace\negthickspace\negthickspace\negthickspace\negthickspace\left\{ \begin{array}{l}
\begin{array}{ccc}
\delta_{\alpha}^{\pm} & \negthickspace\negthickspace\negthickspace= & \negthickspace\negthickspace\negthickspace\{1,0\}\\
\alpha^{\pm} & \negthickspace\negthickspace\negthickspace\leq & \negthickspace\negthickspace\negthickspace C\delta_{\alpha}^{\pm}
\end{array}\\
-P_{G}^{\star}+P_{G}^{\max}\leq C(1-\delta_{\alpha}^{+})\\
\;\:\; P_{G}^{\star}-P_{G}^{\min}\leq C(1-\delta_{\alpha}^{-})
\end{array}\right.\\
 & \negthickspace\negthickspace\negthickspace\negthickspace\negthickspace\negthickspace\negthickspace\negthickspace\negthickspace\left\{ \begin{array}{l}
\begin{array}{ccc}
\delta_{\beta} & \negthickspace\negthickspace\negthickspace= & \negthickspace\negthickspace\negthickspace\{1,0\}\\
\beta & \negthickspace\negthickspace\negthickspace\leq & \negthickspace\negthickspace\negthickspace C\delta_{\beta}
\end{array}\\
\;\; R^{\star}\leq C(1-\delta_{\beta})
\end{array}\right.\label{eq:LinearComplementary2}
\end{align}
where $\delta_{\lambda}^{\pm}$, $\delta_{\alpha}^{\pm}$ and $\delta_{\beta}$
are binary variables and $C$ is a large constant.

\section{Simulation Results \label{sec:Simulation-Results}}

In this section, we run the optimization problem defined in Sec. \ref{sec:optimization-problem}
on the IEEE RTS-24-bus system to find an optimal attack vector $c$. Subsequently, we use this attack
vector $c$ to simulate an AC attack described in Sec. \ref{sub:AC-attack}
and given by \eqref{eq:AC_attack} against a non-linear system model
involving AC state estimation and ACOPF. AC power flow, AC state estimation,
and ACOPF are implemented with MATPOWER toolbox in MATLAB. For the
optimization problem, we use CPLEX as the solver.

\subsection{Solution for the optimization problem\label{sub:Solution-for-the}}

We highlight results of two scenarios for the RTS-24-bus system: one
with original rating and one with reduced rating. The one with original
rating represents a system without congestion prior to attack and
the one with reduced rating represent a congested system.

Second, we define an attack as \emph{feasible} if the resulting change
in power flow is more than $1\%$ of the power flow value prior to
the attack. This is to distinguish the cases with no or minor changes
on target branch power flow $P_{l}$ after attack from those with
large changes. We furthermore define a feasible attack to be \emph{successful}
if the target branch is overloaded after attack. We choose $\gamma$
to be $1\%$ of the original power flow value of the target branch.

\begin{figure}[tbh]
\includegraphics{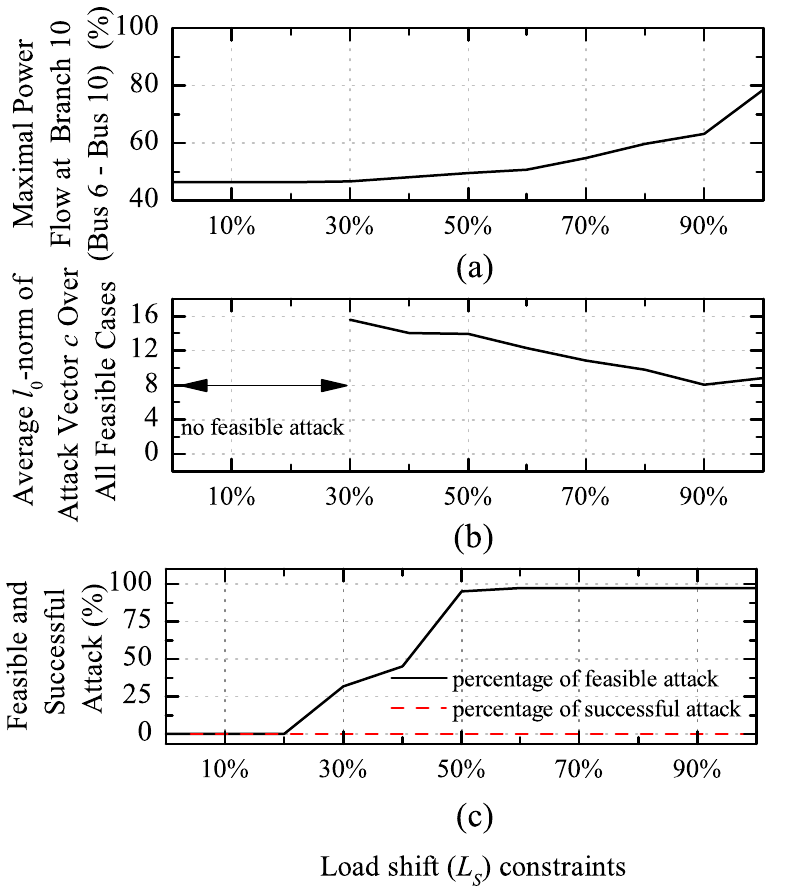}
\caption{Statistic summary of 38 attack scenarios for the omnipotent attacker
with the non-congested system. \label{Fig:table1}}
\end{figure}

\begin{figure}[tbh]
\includegraphics{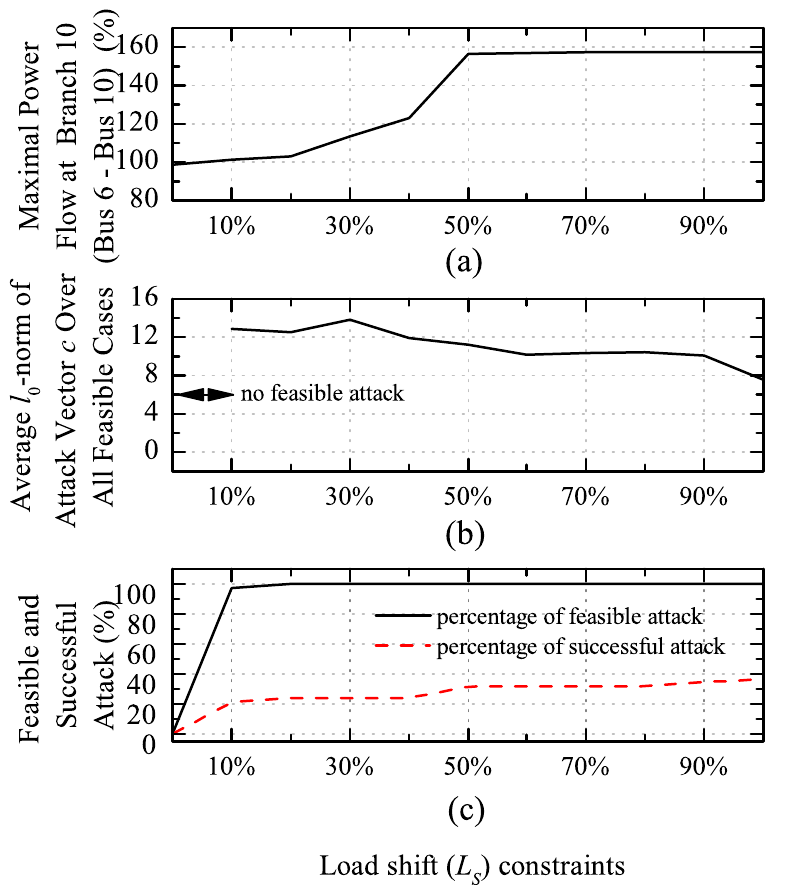}
\caption{Statistic summary of 38 attack scenarios for the omnipotent attacker
with the congested system.\label{Fig:table2} }

\end{figure}

Figs. \ref{Fig:table1} and \ref{Fig:table2} illustrate relevant
statistics for the non-congested and congested systems, respectively,
when the $N_{1}$ constraint is set to be infinite. That is, the attacker
has control over all measurements of the system and can change as
many measurements as it wishes. The congested system is modeled with
all branch ratings decreased by $50\%$. There are three subplots
in both Figs. \ref{Fig:table1} and \ref{Fig:table2}. Subplot (a)
shows the maximal power flow on branch 10 (based on our observation,
this is the attack with the maximal power flow, i.e., the worst-case
attack); subplot (b) shows the average $l_{0}$-norm of attack vector
$c$ over all feasible cases; and subplot (c) shows the percentage
of feasible and successful attacks. 

For both non-congested and congested scenarios, we observe that the
maximal power flow increases as $L_{S}$ constraint relaxes in Figs.
\ref{Fig:table1}(a) and \ref{Fig:table2}(a). In Fig. \ref{Fig:table2}(a),
we observe a plateau after $L_{S}>50\%$. It is due to the generator
location and capacity limitation and the fact that the line flow on
branch 10 cannot be increased anymore. From Figs. \ref{Fig:table1}(b)
and \ref{Fig:table2}(b), as $L_{S}$ constraint relaxes, it is easier
to attack the system since the average $l_{0}$-norm decreases and
the attacker needs to change fewer bus states. It is due to the fact, for some cases, that the maximal power is saturated when the $L_S$ constraint relaxes. The attacker effectively concentrates the change of loads on fewer buses with heavy loads therefore changes fewer bus states. 
 From Figs. \ref{Fig:table1}(c)
and \ref{Fig:table2}(c), we observe that the attacker can find more
feasible cases as $L_{S}$ constraint relaxes. Even if the attacker
has full control over the system meters, its influence over the system
is extremely limited by the load shift constraint. For instance, from
Fig. \ref{Fig:table1}(c), when $L_{S}=20\%$, the attacker cannot
find any feasible attacks while the attacker can find 12 feasible
attacks when $L_{S}=30\%$.

Comparing Figs. \ref{Fig:table1} and \ref{Fig:table2}, the congested
system is more vulnerable to our FDI attack. For a non-congested system,
from Fig. \ref{Fig:table1}(c), the attacker cannot generate any successful
attack. On the other hand, in Fig. \ref{Fig:table2}(c), the feasible
and successful attack percentage increases as $L_{S}$ constraint
increases for the congested system. This is expected because the RTS-24-bus
system has redundant transmission capacity for reliability reasons
and reducing all the line ratings proportionally will create a more
stressed system. In conclusion, a congested system
is naturally favored by the attacker. Thus, for the rest of the simulation,
we only consider the congested system to illustrate the attack consequences. 

\begin{figure}[h]
\includegraphics{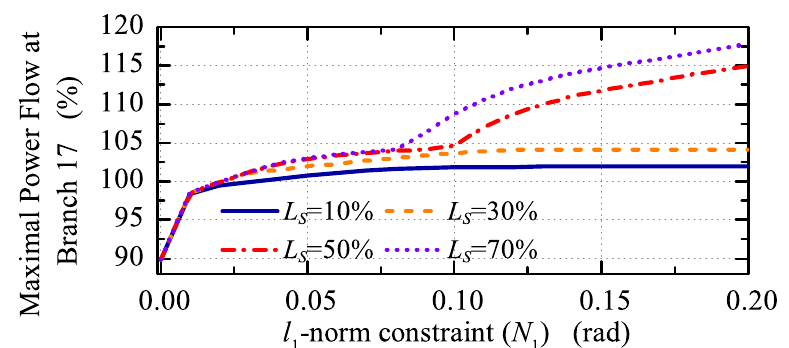}
\caption{The maximal power flow v.s. the $l_{1}$ -norm constraints with different
load shift tolerance at target branch 17 (bus 10-- bus 12).\label{fig:Branch 17 _Pmx} }
\end{figure}

Now we discuss the $l_{1}$-norm constraint. To understand the effect
of the sparsity constraint, we fix the $L_{S}$ constraint and solve the proposed optimization problem for different $l_{1}$-norm
constraint ($N_{1})$ and for all target branches. In Fig. \ref{fig:Branch 17 _Pmx},
the maximal power flow on the target branch is plotted as a function
of the $l_{1}$-norm constraint for a successful attack on target
branch 17. The kink in Fig. \ref{fig:Branch 17 _Pmx} represents point of which the attack is large enough to cause a different set of generators to be dispatched.

\begin{figure}[tbh]
\includegraphics{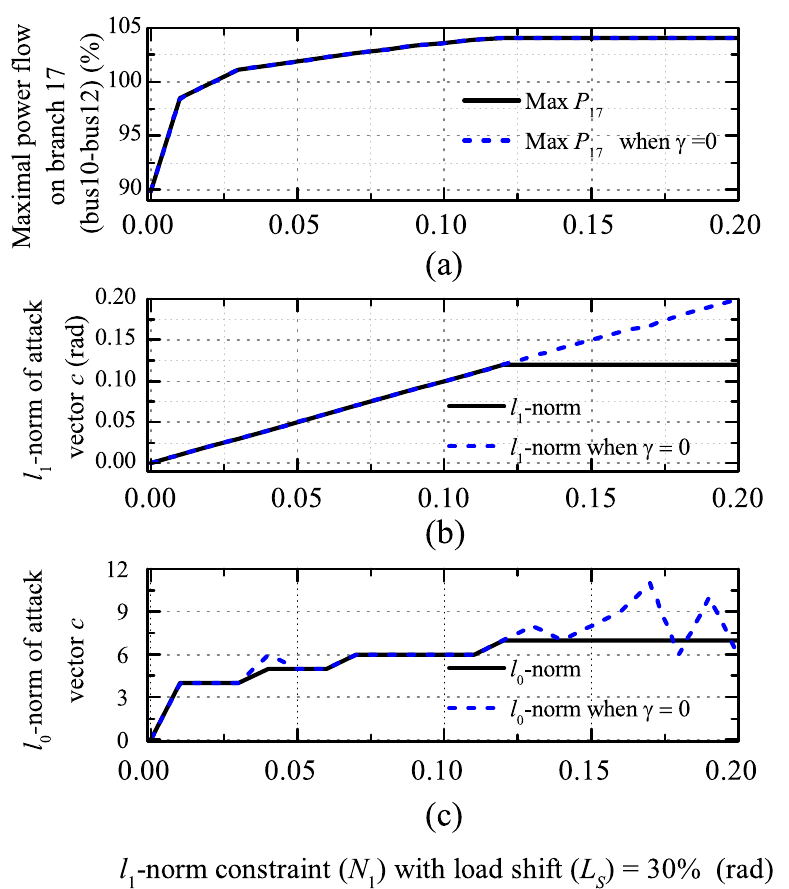}
\caption{The $l_{1}$-norm and $l_{0}$-norm of solved attack vector $c$ v.s.
the $l_{1}$-norm constraint ($N_{1}$) when load shift ($L_{S}$)
is limited by 30\%; target branch 17 (bus 10-- bus 12).\label{fig:Branch17_L0L1}}
\end{figure}

Fig. \ref{fig:Branch17_L0L1} illustrates the effect of the $\gamma$
term in the objective function of our optimization problem for target
branch 17. There are three sub-plots  illustrating
the following as a function of the $l_{1}$-norm constraint $N_{1}$:
(a) the maximal power flow, (b) the $l_{1}$-norm , and (c) the $l_{0}$-norm
of the attack vector, respectively. In each subplot, we plot two curves,
one with $\gamma$ set to zero and one with the chosen weight of $\gamma$
coefficient. Subplot (a) demonstrates that the $\gamma$ term does
not decrease the resulting maximal power flow at all. Subplot (b) shows that
once the maximal power flow saturates,  introducing the $\gamma$ term causes
the optimization problem to find the smallest attack vector in $l_{1}$-norm.
This result in a stabilization of the $l_{0}$-norm as shown in subplot
(c) in contrast to the $\gamma=0$ case.

\subsection{Attack consequences for a non-linear model\label{sub:Simulation-of-consequences}}

\begin{figure}[tbh]
\subfloat[\label{fig:AC_branch17} Comparison of DC optimization solution and
AC maximal active/apparent power flow on target branch 17]{
\includegraphics{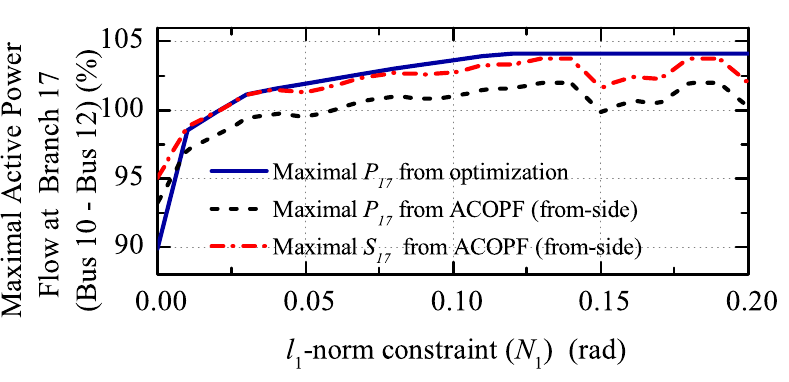}
}

\subfloat[\label{fig:AC_other_branches}Power flow on branch 12, 23, and 28]{
\includegraphics{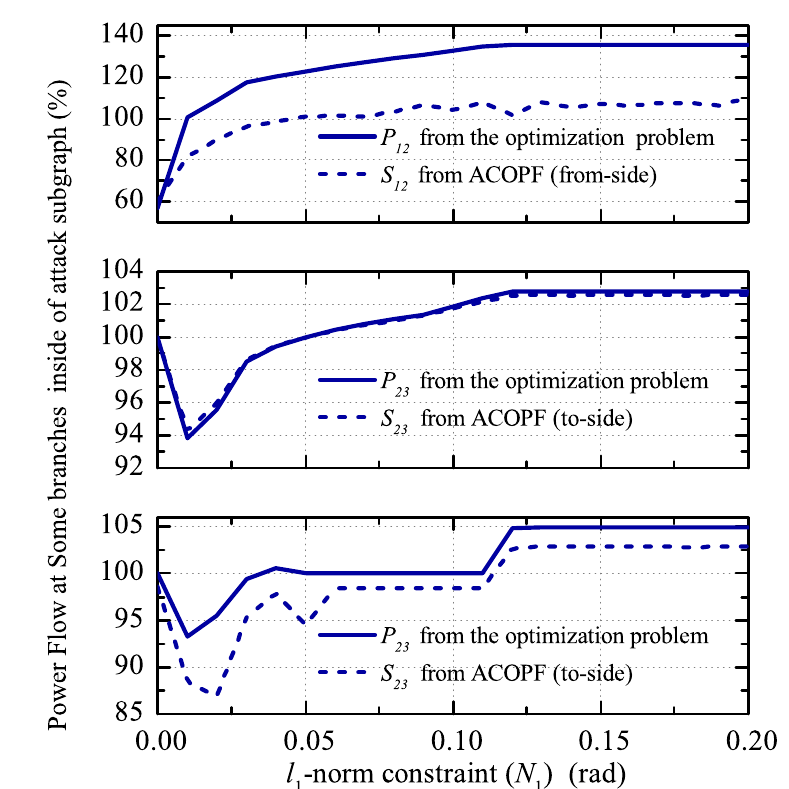}

}\caption{Attack simulation on AC system model\label{fig:AC attack }}
\label{Fig:AC_DC_Compare}
\end{figure}

We now use the attack vector from the optimization problem to perform
the AC attack described in Sec. \ref{sub:AC-attack}. If the attacker
keeps injecting false data, the attack as well as the overload on
the branches will be sustained until the system configuration changes. 

In this subsection, we assume a system with a complete set of measurements,
i.e., both active and reactive power flows are measured at two ends
of each branch and both active and reactive injection are measured
at each load bus, which makes 186 measurements in total. Measurement error, as in \eqref{Measurement_model}, is assumed to have zero mean and variance $10^{-4}$. During the
simulation, we assume the physical load is unchanged. Note that, to
make the system congested, all ratings of the branches are decreased
by $50\%$. However, because of the existence of the reactive power,
the convergence of ACOPF is not guaranteed. Thus, in order to compare
AC and DC attack, certain ratings of branches in ACOPF have to be
relaxed manually. We solve the optimization problem for target branch
17 and $L_{S}=30\%$. Fig. \ref{fig:AC attack }(a) compares the maximal
power flow from the optimization problem (recall: DCOPF used in the
optimization problem) with the physical power flow (active and apparent
power) in the non-linear system after attack. In this scenario, the
rating of branch 10 (bus 6--bus 10) are relaxed to 145 MVA. 
Note that in the absent of the attack, i.e., $N_{1}=0$, the power
flow for AC and DC OPFs result in sightly different power flow, however,
as the attacker size is increased, the power flows closely track each other. In particularly, 
the AC attack successfully overloads branch 17.
Branches as 12, 23, and 28 are also overloaded even though the attacker has  not targeted
on them, as shown in Fig. \ref{fig:AC attack }(b). Since branch 23
and 28 are congested prior to attack, once the generation is redispatched
as a result of the attack, the power flow on these branches will change
and in some cases it leads to overloads. Branch 12, while not congested prior to attack, suffers an overload due to the fact that it lies in the path of  power delivering to branch 17.

\section{Conclusions and future work \label{sec:Conclusions-and-future}}

This paper analyzed the physical consequences of false data injection attacks on power system state estimation. An attack framework was introduced in which the attacker matches the non-linear AC system characteristics by implementing local AC state estimation to a small number of measurements. Subsequently, a linear optimization problem was formulated to find the worst-case line overload attack. Numerical simulation was performed to test the resulting attacks on the IEEE-RTS-24-bus system. We found that, aside from the size of the attack subgraph, the constraint that an attack not cause significant observed load shift at the control center significantly impacts the attacker's ability to overload a branch. Still, there exists attacks with mild load shift that cause overloads.

Extensions include attacks targeted to overload multiple lines; this was an inadvertent side effect of our attacks, but a more targeted effort may cause more extreme damage or even cascading outages. Secondly, the linear optimization problem may be extended to a more accurate non-linear problem. Finally, using accurate load statistics to detect abnormal load patterns caused by FDI attacks could further restrict the space of undetectable attacks.

\bibliographystyle{IEEEtran}
\bibliography{Ref}

\end{document}